\documentclass{article}
\usepackage{amsmath,amssymb,amsfonts,url}
\usepackage{graphicx}
\usepackage{color}
\usepackage{subfig}
\newcommand{\mb}{\mathbf}
\usepackage[margin=1in]{geometry}

\title{Cover Song Synthesis by Analogy}
\date{}

\author
 {Christopher J. Tralie \\ Duke University Department of Mathematics \\ \texttt{ctralie@alumni.princeton.edu }}

\begin{document}

\maketitle
\begin{abstract}
﻿In this work, we pose and address the following ``cover song analogies'' problem: given a song A by artist 1 and a cover song A' of this song by artist 2, and given a different song B by artist 1, synthesize a song B' which is a cover of B in the style of artist 2. Normally, such a polyphonic style transfer problem would be quite challenging, but we show how the cover songs example constrains the problem, making it easier to solve. First, we extract the longest common beat-synchronous subsequence between A and A', and we time stretch the corresponding beat intervals in A' so that they align with A. We then derive a version of joint 2D convolutional NMF, which we apply to the constant-Q spectrograms of the synchronized segments to learn a translation dictionary of sound templates from A to A'. Finally, we apply the learned templates as filters to the song B, and we mash up the translated filtered components into the synthesized song B' using audio mosaicing. We showcase our algorithm on several examples, including a synthesized cover version of Michael Jackson's ``Bad'' by Alien Ant Farm, learned from the latter's ``Smooth Criminal'' cover.
\end{abstract}
\section{Introduction}\label{sec:introduction}

\begin{figure}[!h]
	\centering	\includegraphics[width=0.8\columnwidth]{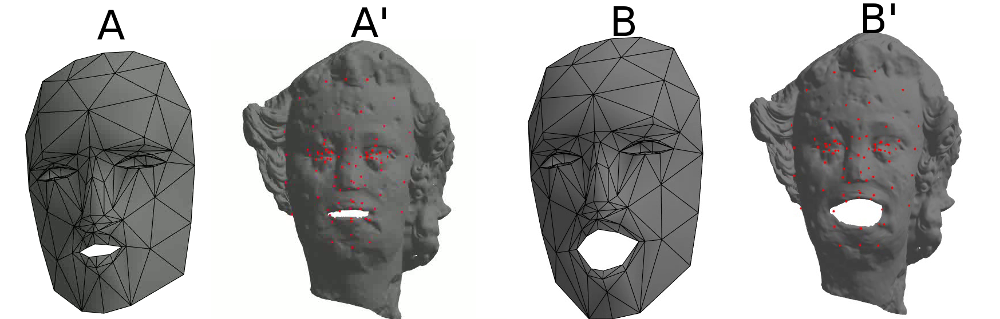}
	\caption{A demonstration of the ``shape analogies'' technique in \cite{sumner2004deformation}.  In the language of our work, the statue head with a neutral expression ($A'$) is a ``cover'' of the low resolution mesh $A$ with a neutral expression, and this is used to synthesize the surprised statue ``cover'' face $B'$ from a low resolution surprised face mesh $B$.}
    \label{fig:SumnerPopovic}
\end{figure}

\begin{figure}[]
	\centering	\includegraphics[width=0.9\columnwidth]{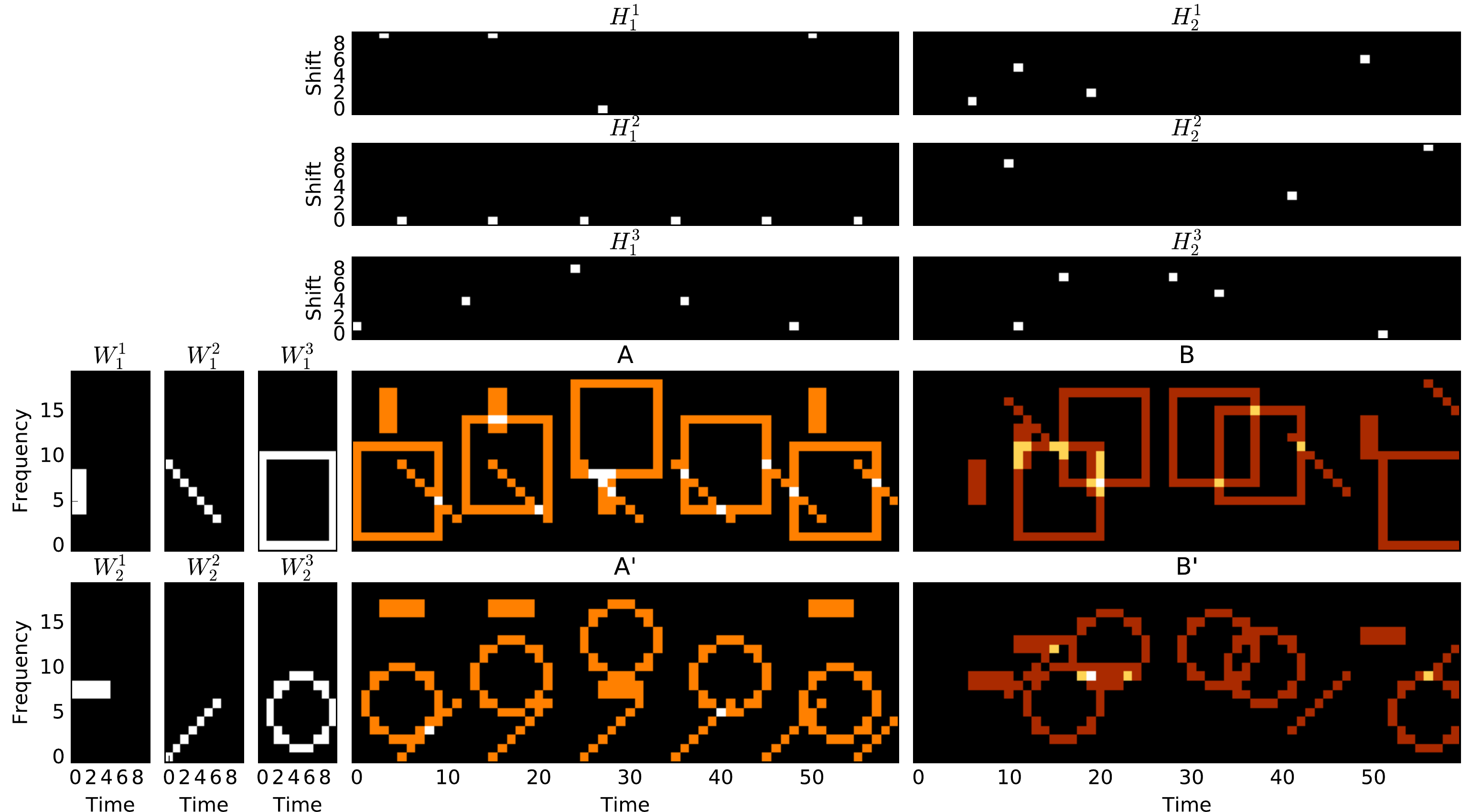}
	\caption{An ideal cartoon example of our joint 2DNMF and filtering process, where $M = 20$, $N=60$, $T=10$, $F=10$, and $K=3$.  In this example, vertical time-frequency blocks are ``covered'' by horizontal blocks, diagonal lines with negative slopes are covered by diagonal lines with positive slopes, and squares are covered by circles.  When presented with a new song $B$, our goal is to synthesize a song $B'$ whose CQT is shown in the lower right green box.}
    \label{fig:SyntheticJoint}
\end{figure}

The rock group Alien Ant Farm has a famous cover of Michael Jackson's ``Smooth Criminal'' which is faithful to but stylistically unique from the original song.  However, to our knowledge, they never released a cover of any other Michael Jackson songs.  What if we instead wanted to know how they would have covered Michael Jackson's ``Bad''?  That is, we seek a song which is identifiable as MJ's ``Bad,'' but which also sounds as if it's in Alien Ant Farm's {\em style}, including timbral characteristics, relative tempo, and instrument types.  

In general, multimedia style transfer is a challenging task in computer aided creativity applications.  When an example of the stylistic transformation is available, as in the ``Smooth Criminal'' example above, this problem can be phrased in the language of analogies; given an object $A$ and a differently stylized version of this object $A'$, and given an object $B$ in the style of $A$, synthesize an object $B'$ which has the properties of $B$ but the style of $A'$.  One of the earliest works using this vocabulary is the ``image analogies'' work \cite{hertzmann2001image}, which showed it was possible to transfer both linear filters (e.g. blurring, embossing) and nonlinear filters (e.g. watercolors) in the same simple framework.  More recent work with convolutional networks has shown even better results for images \cite{gatys2015neural}.  There has also been some work on ``shape analogies'' for 3D meshes \cite{sumner2004deformation}, in which nonrigid deformations between triangle meshes $A$ and $A'$ are used to induce a corresponding deformation $B'$ from an object $B$, which can be used for motion transfer (Figure~\ref{fig:SumnerPopovic}).  

In the audio realm, most style transfer works are based on mashing up sounds from examples in a target style using ``audio mosaicing,'' usually after manually specifying some desired path through a space of sound grains \cite{schwarz2008principles}.  A more automated audio moscaicing technique, known as ``audio analogies'' \cite{simon2005audio}, uses correspondences between a MIDI score and audio to drive concatenated synthesis, which leads to impressive results on monophonic audio, such as stylized synthesis of jazz recordings of a trumpet.  More recently, this has evolved into the audio to musical audio setting with audio ``musaicing,'' in which the timbre of an audio source is transferred onto a target by means of a modified NMF algorithm \cite{driedger2015let}, such as bees buzzing The Beatles' ``Let It Be.''  A slightly closer step to the polyphonic (multi source) musical audio to musical audio case has been shown to work for drum audio cross-synthesis with the aid of a musical score \cite{DittmarMueller16_DrumSeparation_IEEE-TASLP}, and some very recent initial work has extended this to the general musical audio to musical audio case \cite{foroughmand2017} using 2D nonnegative matrix factorization, though this still remains open.  Finally, there is some recent work on converting polyphonic audio of guitar songs to musical scores of varying difficulties so users can play their own covers \cite{ariga2017guitarcover}.

In this work, we constrain the polyphonic musical audio to musical audio style transfer problem by using {\em cover song} pairs, or songs which are the same but in a different style, and which act as our ground truth $A$ and $A'$ examples in the analogies framework.  Since small scale automatic cover song identification has matured in recent years \cite{serra2009cross, silvasimple, Chen2017CSFusion, tralie2017cover}, we can accurately synchronize $A$ and $A'$ (Section~\ref{sec:Alignment}), even if they are in very different styles.  Once they are synchronized, the problem becomes more straightforward, as we can blindly factorize $A$ and $A'$ into different instruments which are in correspondence, turning the problem into a series of monophonic style transfer problems.  To do this, we perform NMF2D factorization of $A$ and $A'$ (Section~\ref{sec:NMF2D}).  We then filter $B$ by the learned NMF templates and mash up audio grains to create $B'$, using the aforementioned ``musaicing'' techniques \cite{driedger2015let} (Section~\ref{sec:musaicing}).  We demonstrate our techniques on snippets of $A$, $A'$, and $B$ which are about 20 seconds long, and we show qualitatively how the instruments of $A'$ transfer onto the music of $B$ in the final result $B'$ (Section~\ref{sec:experiments}).


\section{Algorithm Details}

In this section, we will describe the steps of our algorithm in more detail.\footnote{Note that all audio is mono and sampled at 22050hz.}

\subsection{Cover Song Alignment And Synchronization}
\label{sec:Alignment}

\begin{figure}[!h]
	\centering	\includegraphics[width=0.8\columnwidth]{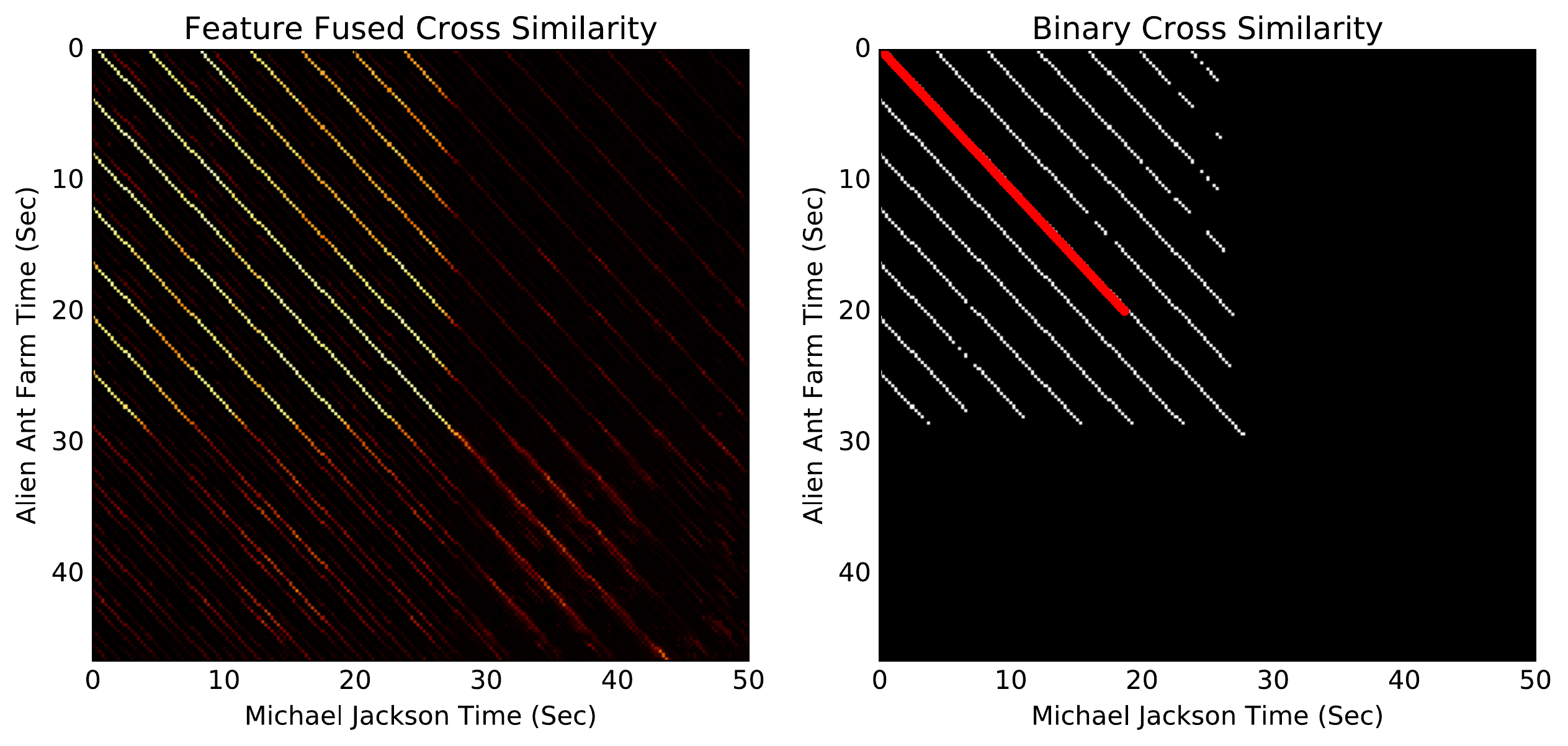}
	\caption{An example feature fused cross-similarity matrix $D$ for the first 80 beats of Michael Jackson's ``Smooth Criminal,'' compared to the cover version by Alien Ant Farm.  We threshold the matrix and extract 20 seconds of the longest common subsequence, as measured by Smith Waterman.  The alignment path is shown in red.}
    \label{fig:MJSynced}
\end{figure}

As with the original image analogies algorithm \cite{hertzmann2001image}, we find it helpful if $A$ and $A'$ are in direct correspondence at the sample level.  Since cover song pairs generally differ in tempo, we need to align them first.  To accomplish this, we draw upon the state of the art ``early fusion'' cover song alignment technique presented by the authors of \cite{tralie2017cover}.  Briefly, we extract beat onsets for $A$ and $A'$ using either a simple dynamic programming beat tracker \cite{ellis2007beat} or slower but more accurate RNN + Bayesian beat trackers \cite{krebs2015efficient}, depending on the complexity of the audio.  We then compute beat-synchronous sliding window HPCP and MFCC features, and we fuse them using similarity network fusion \cite{wang2012unsupervised, wang2014similarity}.  The result is a $M \times N$ cross-similarity matrix $D$, where $M$ is the number of beats in $A$ and $N$ is the number of beats in $A'$, and $D_{ij}$ is directly proportional to the similarity between beat $i$ of $A$ and beat $j$ in $A'$.  Please refer to \cite{tralie2017cover} for more details.

Once we have the matrix $D$, we can then extract an alignment between $A$ and $A'$ by performing Smith Waterman \cite{smith1981identification} on a binary thresholded version of $D$, as in \cite{tralie2017cover}.  We make one crucial modification, however.  To allow for more permissive alignments with missing beats for identification purposes, the original cover songs algorithm creates a binary thresholded version of $D$ using $10 \%$ mutual binary nearest neighbors.  On the other hand, in this application, we seek shorter snippets from each song which are as well aligned as possible.  Therefore, we create a stricter binary thresholded version $B$, where $B_{ij} = 1$ only if it is in the top $3\sqrt{MN}$ distances over all $MN$ distances in $D$.  This means that many rows of $B_{ij}$ will be all zeros, but we will hone in on the best matching segments.  Figure~\ref{fig:MJSynced} shows such a thresholding of the cross-similarity matrix for two versions of the song ``Smooth Criminal,'' which is an example we will use throughout this section.  Once $B$ has been computed, we compute a $X$-length alignment path $P$ by back-tracing through the Smith Waterman alignment matrix, as shown in Figure~\ref{fig:MJSynced}.

Let the beat onset times for $A$ in the path $P$ be $t_1, t_2, ..., t_X$ and the beat times for $A'$ be $s_1, s_2, ..., s_X$.  We use the rubberband library \cite{cannam2012rubber} to time stretch $A'$ beat by beat, so that interval $[s_i, s_i+1]$ is stretched by a factor $(t_{i+1}-t_i) / (s_{i+1}-s_i)$.  The result is a snippet of $A'$ which is the same length as the corresponding snippet in $A$.  Henceforth, we will abuse notation and refer to these snippets as $A$ and $A'$.  We also extract a smaller snippet from $B$ of the same length for reasons of memory efficiency, which we will henceforth refer to as $B$.

\subsection{NMF2D for Joint Blind Factorization / Filtering}
\label{sec:NMF2D}

\begin{figure*}[]
	\centering	\includegraphics[width=\textwidth]{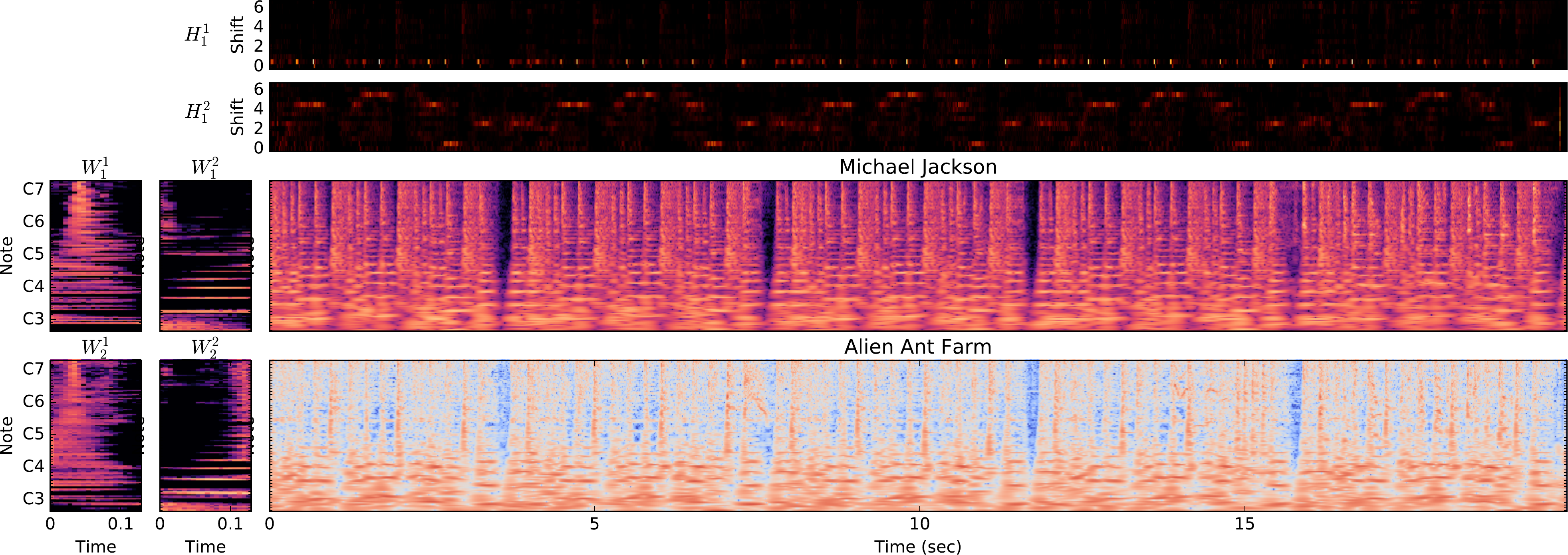}
	\caption{Joint 2DNMF on the magnitude CQTs of ``Smooth Criminal'' by Michael Jackson and Alien Ant Farm, with $F=14$ (7 halfsteps at 24 bins per octave), $T=20$ (130ms in time), and $K=2$ components.  In this case, $W_1^1$ and $W_2^1$ hold percussive components, and $W_1^2$ and $W_2^2$ hold guitar components.}
    \label{fig:MJJoint}
\end{figure*}

Once we have synchronized the snippets $A$ and $A'$, we blindly factorize and filter them into $K$ corresponding tracks $A_1, A_2, ..., A_K$ and $A'_1, A'_2, ..., A'_K$.  The goal is for each track to contain a different instrument.  For instance, $A_1$ may contain an acoustic guitar, which is covered by an electric guitar contained in $A'_1$, and $A_2$ may contain drums which are covered by a drum machine in $A'_2$.  This will make it possible to reduce the synthesis problem to $K$ independent monophonic analogy problems in Section~\ref{sec:musaicing}.

To accomplish this, the main tool we use is 2D convolutional nonnegative matrix factorization (2DNMF) \cite{schmidt2006nonnegative}.  We apply this algorithm to the magnitude of the constant-Q transforms (CQTs) \cite{brown1991calculation} $\mb{C_A}$, $\mb{C_{A'}}$ and $\mb{C_{B}}$ of $A$, $A'$, and $B$, respectively.  Intuitively, we seek $K$ time-frequency templates in $A$ and $A'$ which represent small, characteristic snippets of the $K$ instruments we believe to be present in the audio.  We then approximate the magnitude constant-Q transform of $A$ and $A'$ by convolving these templates in both time and frequency.  The constant-Q transform is necessary in this framework, since a pitch shift can be approximated by a {\em linear} shift of all CQT bins by the same amount.  Frequency shifts can then be represented as convolutions in the vertical direction, which is not possible with the ordinary STFT.  Though it is more complicated, 2DNMF is a more compact representation than 1D convolutional NMF (in time only), in which it is necessary to store a different template for each pitch shift of each instrument.  We note that pitch shifts in real instruments are more complicated than shifting all frequency bins by the same perceptual amount \cite{nakamura2016shifted}, but the basic version of 2DNMF is fine for our purposes.

More concretely, define a $K$-component, $F$-frequency, $T$-time lag 2D convolutional NMF decomposition for a matrix $\mb{X} \in \mathbb{R}^{M \times N}$ as follows

\begin{equation}
\label{eq:NMF2D}
\mb{X} \approx \mb{\Lambda_{W,H}} = \sum_{\tau=1}^T \sum_{\phi=1}^F \overset{\downarrow \phi}{\mb{W}^{\tau}} \overset{\rightarrow \tau}{\mb{H}^{\phi}}
\end{equation}

where $\mb{W^{\tau}} \in \mathbb{R}^{M \times K}$ and $\mb{H^{\phi}} \in \mathbb{R}^{K \times N}$ store a $\tau$-shifted time template and $\phi$-frequency shifted coefficients, respectively.  By $\overset{\downarrow \phi}{A}$, we mean down shift the rows of $A$ by $\phi$, so that row $i$ of $\overset{\downarrow \phi}{A}$ is row $i-\phi$ of $A$, and the first $\phi$ rows of $\overset{\downarrow \phi}{A}$ are all zeros.  And $\overset{\leftarrow \tau}{A}$ means left-shift $A$, so that column $j$ of $\overset{\leftarrow \tau}{A}$ is column $j-\tau$ of $A$, and the first $\tau$ columns of $\overset{\leftarrow \tau}{A}$ are all zeros.

In our problem, we define an extension of 2DNMF to {\em jointly} factorize the 2 songs, $A$ and $A'$, which each have $M$ CQT coefficients.  In particular, given matrices $\mb{C_A}, \mb{C_{A'}} \in \mathbb{C}^{M \times N_1}$ representing the complex CQT frames in each song over time, we seek $\mb{W_1^{\tau}}, \mb{W_2^{\tau}} \in \mathbb{R}^{M \times K}$ and $\mb{H_1^{\phi}} \in \mathbb{R}^{K \times N_1}$ that minimize the sum of the Kullback-Leibler divergences between the magnitude CQT coefficients and the convolutions:

\begin{equation}
\label{eq:error}
D(\mb{|C_A|} \; || \; \mb{\Lambda_{W_1, H_1}}) + D(\mb{|C_{A'}|} \; || \; \mb{\Lambda_{W_2, H_1}})
\end{equation}

where the Kullback-Leibler divergence $D(X || Y)$ is defined as

\begin{equation}
D(\mb{X} || \mb{Y}) = \sum_{i} \sum_{j} \mb{X}_{i, j} \log \frac{ {\mb{X}_{i, j}} }{\mb{Y}_{i, j}} - \mb{X}_{i, j} + \mb{Y}_{i, j}
\end{equation}

That is, we {\em share} $\mb{H_1}$ between the factorizations of $|\mb{C_A}|$ and $|\mb{C_{A'}}|$ so that we can discover shared structure between the covers.  Following similar computations to those of ordinary 2DNMF \cite{schmidt2006nonnegative}, it can be shown that Equation~\ref{eq:error} is non-decreasing under the alternating update rules:

\begin{equation}
\label{eq:updateW1}
\mb{W_1^{\tau}} \gets \mb{W_1^{\tau}} \odot
    \frac{
    \sum_{\phi=1}^F \overset{\uparrow \phi}{ \left( \frac{ \mb{|C_A|} }{ \mb{\Lambda_{W_1, H_1}} } \right)}
    \overset{\rightarrow \tau}{\mb{H_1^{\phi}}}^T
    }
    {\sum_{\phi=1}^F
    \mb{1} \cdot \overset{\rightarrow \tau}{\mb{H_1^{\phi}}}^T }
\end{equation}

\begin{equation}
\label{eq:updateW2}
\mb{W_2^{\tau}} \gets \mb{W_2^{\tau}} \odot
    \frac{
    \sum_{\phi=1}^F \overset{\uparrow \phi}{ \left( \frac{ \mb{|C_{A'}|} }{ \mb{\Lambda_{W_2, H_1}} } \right)}
    \overset{\rightarrow \tau}{\mb{H_1^{\phi}}}^T
    }
    {\sum_{\phi=1}^F
    \mb{1} \cdot \overset{\rightarrow \tau}{\mb{H_1^{\phi}}}^T }
\end{equation}

\begin{equation}
\label{eq:updateH}
\mb{H_1^{\phi}} \gets \mb{H_1^{\phi}} \odot
\left( \frac{ \sum_{\tau=1}^T \overset{\downarrow \phi}{\mb{W_1^{\tau}}}^T \overset{\leftarrow \tau}{ \left( \frac{\mb{|C_A|}}{\mb{\Lambda_{W_1, H_1}}} \right) } + \overset{\downarrow \phi}{\mb{W_2^{\tau}}}^T  \overset{\leftarrow \tau}{ \left( \frac{\mb{|C_{A'}|}}{\mb{\Lambda_{W_2, H_1}}} \right) }  }
{ \sum_{\tau=1}^T  \overset{\downarrow \phi}{\mb{W_1^{\tau}}}^T \overset{\leftarrow \tau}{\mb{1}} +  \overset{\downarrow \phi}{\mb{W_2^{\tau}}}^T \overset{\leftarrow \tau}{\mb{1}}}
\right)
\end{equation}

where $\mb{1}$ is a column vector of all 1s of appropriate dimension.  We need an invertible CQT to go back to audio templates, so we use the non-stationary Gabor Transform (NSGT) implementation of the CQT \cite{velasco2011constructing} to compute $\mb{C_A}, \mb{C_{A'}}$, and $\mb{C_B}$.  We use 24 bins per octave between 50hz and 11.7kHz, for a total of 189 CQT bins.  We also use $F = 14$ in most of our examples, allowing 7 halfstep shifts, and we use $T = 20$ on temporally downsampled CQTs to cover a timespan of 130 milliseconds.  Finally, we iterate through Equations~\ref{eq:updateW1},~\ref{eq:updateW2},and~\ref{eq:updateH} in sequence 300 times.  

Note that a naive implementation of the above equations can be very computationally intensive.  To ameliorate this, we implemented GPU versions of Equations~\ref{eq:NMF2D},~\ref{eq:updateW1},~\ref{eq:updateW2},and~\ref{eq:updateH}.  Equation~\ref{eq:NMF2D} in particular is well-suited for a parallel implementation, as the shifted convolutional blocks overlap each other heavily and can be carefully offloaded into shared memory to exploit this\footnote{In the interest of space, we omit more details of our GPU-based NMFD in this paper, but a documented implementation can be found at \url{https://github.com/ctralie/CoverSongSynthesis/}, and we plan to release more details in a companion paper later.}.  In practice, we witnessed a 30x speedup of our GPU implementation over our CPU implementation for 20 second audio clips for $A$, $A'$, and $B$.

Figure~\ref{fig:SyntheticJoint} shows a synthetic example with an exact solution, and Figure~\ref{fig:MJJoint} shows a local min which is the result of running Equations~\ref{eq:updateW1},~\ref{eq:updateW2},and~\ref{eq:updateH} on real audio from the ``Smooth Criminal'' example.  It is evident from $\mb{H_1}$ that the first component is percussive (activations at regular intervals in $\mb{H_1^1}$, and no pitch shifts), while the second component corresponds to the guitar melody ($\mb{H_1^2}$ appears like a ``musical score'' of sorts).  Furthermore, $\mb{W_1^1}$ and $\mb{W_2^1}$ have broadband frequency content consistent with percussive events, while $\mb{W_1^2}$ and $\mb{W_2^2}$ have visible harmonics consistent with vibrating strings.  Note that we generally use more than $K=2$, which allows finer granularity than harmonic/percussive, but even for $K=2$ in this example, we observe qualitatively better separation than off-the-shelf harmonic/percussive separation algorithms \cite{driedger2014extending}.

Once we have $\mb{W_1}$, $\mb{W_2}$, and $\mb{H_1}$, we can recover the audio templates $A_1, A_2, ..., A_K$ and $A'_1, A'_2, ..., A'_K$ by using the components of $\mb{W_1}$ and $\mb{W_2}$ as filters.  First, define $\mb{\Lambda_{W, H, k}}$ as

\begin{equation}
\mb{\Lambda_{W,H, k}} = \sum_{\tau=1}^T \sum_{\phi=1}^F \overset{\downarrow \phi}{\mb{W^k}^{\tau}} \overset{\rightarrow \tau}{\mb{H_k}^{\phi}}
\end{equation}

where $\mb{W^k}$ is the $k^{\text{th}}$ column of $\mb{W}$ and $\mb{H_k}$ is the $k^{\text{th}}$ row of $\mb{H}$.  Now, define the filtered CQTs by using soft masks derived from $\mb{W_1}$ and $\mb{H_1}$:

\begin{equation}
\mb{C_{A_k}} = \mb{C_A} \odot \left( \frac{\mb{\Lambda_{W_1, H_1, k}^p}}{\sum_{m=1}^K \mb{\Lambda_{W_1, H_1, m}^p}} \right)
\end{equation}

\begin{equation}
\mb{C_{A'_k}} = \mb{C_{A'}} \odot \left( \frac{\mb{\Lambda_{W_2, H_1, k}^p}}{\sum_{m=1}^K \mb{\Lambda_{W_2, H_1, m}^p}} \right)
\end{equation}

where $p$ is some positive integer applied element-wise (we choose $p = 2$), and the above multiplications and divisions are also applied element-wise.  It is now possible to invert the CQTs to uncover the audio templates, using the inverse NSGT \cite{velasco2011constructing}.  Thus, if the separation was good, we are left with $K$ independent monophonic pairs of sounds between $A$ and $A'$.

In addition to inverting the sounds after these masks are applied, we can also listen to the components of $\mb{W_1}$ and $\mb{W_2}$ themselves to gain insight into how they are behaving as filters.  Since $\mb{W_1}$ and $\mb{W_2}$ are magnitude only, we apply the Griffin Lim algorithm \cite{griffin1984signal} to perform phase retrieval, and then we invert them as before to obtain 130 millisecond sounds for each $k$.

\subsection{Musaicing And Mixing}
\label{sec:musaicing}

We now describe how to use the corresponding audio templates we learned in Section~\ref{sec:NMF2D} to perform style transfer on a new piece of audio, $B$.  

\subsubsection{Separating Tracks in $B$}
First, we compute the CQT of $B$, $\mb{C_B} \in \mathbb{C}^{M \times N_2}$.  We then represent its magnitude using $\mb{W_1}$ as a basis, so that we filter $B$ into the same set of instruments into which $A$ was separated.  That is, we solve for $\mb{H_2}$ so that $|\mb{C_B}| \approx \mb{\Lambda_{W_1, H_2}}$.  This can be performed with ordinary 2DNMF, holding $\mb{W_1}$ fixed; that is, repeating the following update until convergence

\begin{equation}
\mb{H_2^{\phi}} \gets \mb{H_2^{\phi}} \odot
\left( \frac{ \sum_{\tau=1}^T \overset{\downarrow \phi}{\mb{W_1^{\tau}}}^T \overset{\leftarrow \tau}{ \left( \frac{\mb{|C_B|}}{\mb{\Lambda_{W_1, H_2}}} \right) }   }
{ \sum_{\tau=1}^T  \overset{\downarrow \phi}{\mb{W_1^{\tau}}}^T \overset{\leftarrow \tau}{\mb{1}} }
\right)
\end{equation}

As with $A$ and $A'$, we can now filter $B$ into a set of audio tracks $B_1, B_2, ..., B_K$ by first computing filtered CQTs as follows

\begin{equation}
\mb{C_{B_k}} = \mb{C_{B}} \odot \left( \frac{\mb{\Lambda_{W_1, H_2, k}^p}}{\sum_{m=1}^K \mb{\Lambda_{W_1, H_2, m}^p}} \right)
\end{equation}

and then inverting them.

\subsubsection{Constructing $B'$ Track by Track}

At this point, we could use $\mb{H_2}$ and let our cover song CQT magnitudes $| \mb{C_{B'}} | =\mb{\Lambda_{W_2,H_2}} $, followed by Griffin Lim to recover the phase.  However, we have found that the resulting sounds are too ``blurry,'' as they lack all but re-arranged low rank detail from $A'$.  Instead, we choose to draw sound grains from the inverted, filtered tracks from $A'$, which contain all of the detail of the original song.  For this, we first reconstruct each track of $B$ using audio grains from the corresponding tracks in $A$, and then we replace each track with the corresponding track in $B$.  To accomplish this, we apply the audio musaicing technique of Driedger \cite{driedger2015let} to each track in $B$, using source audio $A$.

For computational and memory reasons, we now abandon the CQT and switch to the STFT with hop size $h = 256$ and window size $w = 2048$.  More specifically, let $N_1$ be the number of STFT frames in $A$ and $A'$ and let $N_2$ be the number of STFT frames in $B$.  For each $A_k$, we create an STFT matrix $\mb{S_{A_k}}$ which holds the STFT of $A_k$ concatenated to pitch shifted versions of $A_k$, so that pitches beyond what were in $A$ can be represented, but by the same instruments.  We use $\pm$ 6 halfsteps, so $\mb{S_{A_k}} \in \mathbb{C}^{w \times 13 N_1}$.  We do the same for $A'_k$ to create $\mb{S_{A'_k}} \in \mathbb{C}^{w \times 13 N_1}$.  Finally, we compute the STFT of $B_k$ without any shifts: $\mb{S_{B_k}} \in \mathbb{C}^{w \times N_2}$.

Now, we apply Driedger's technique, using $|\mb{S_{A_k}}|$ as a spectral dictionary to reconstruct $|\mb{S_{B_k}}|$ ($\mb{S_{A_k}}$ is analogous to the buzzing bees in \cite{driedger2015let}).  That is, we seek an $\mb{H_k} \in \mathbb{R}^{13 N_1 \times N_2}$ so that

\begin{equation}
|\mb{S_{B_k}}| \approx |\mb{S_{A_k}}| \mb{H}
\end{equation}

For completeness, we briefly re-state the iterations in Driedger's technique \cite{driedger2015let}.  For $L$ iterations total, at the $\ell^{\text{th}}$ iteration, compute the following 4 updates in order.  First, we restrict the number of repeated activations by filtering in a maximum horizontal neighborhood in time

\begin{equation}
\mb{R_{km}^{(\ell)}} = \left\{  \begin{array}{cc} \mb{H_{km}^{(\ell)}} & \text{if  } \mb{H_{km}^{(\ell)}} = \mb{\mu_{km}^{r,(\ell)}}   \\ \mb{H_{km}^{(\ell)}} (1 - \frac{n+1}{N}) & \text{otherwise} \end{array} \right\}
\end{equation}

where $\mb{\mu_{km}^{r,(\ell)}}$ holds the maximum in a neighborhood $\mb{H_{k, m-r:m+r}}$ for some parameter $r$ (we choose $r = 3$ in our examples).  Next, we restrict the number of simultaneous activations by shrinking all of the values in each column that are less than the top $p$ values in that column:

\begin{equation}
\mb{P_{km}^{(\ell)}} = \left\{  \begin{array}{cc}  \mb{R_{km}^{(\ell)}} & \text{if  } \mb{R_{km}^{(\ell)}} \geq \mb{M^{p (n)}}  \\ \mb{R_{km}^{(\ell)}} (1 - \frac{n+1}{N}) & \text{otherwise} \end{array}  \right\}
\end{equation}

where $\mb{M^{p (\ell)}}$ is a row vector holding the $p^{\text{th}}$ largest value of each column of $\mb{R}^{(\ell)}$ (we choose $p=10$ in our examples).  After this, we promote time-continuous structures by convolving along all of the diagonals of $\mb{H}$:

\begin{equation}
\mb{C_{km}^{(\ell)}} = \sum_{i=-c}^{c} \mb{P^{(\ell)}_{(k+i),(m+i)}}
\end{equation}

We choose $c = 3$ in our examples.  Finally, we perform the ordinary KL-based NMF update:

\begin{equation}
\mb{H^{(\ell+1)}} \gets \mb{H^{(\ell)}} \odot \frac{|\mb{S_{A_k}}|^T \frac{|\mb{S_{B_k}}|}{|\mb{S_{A_k}}| \mb{C^{(\ell)}}} }{ \mb{|S_{A_k}|} \cdot \mb{1} } 
\end{equation}

We perform 100 such iterations ($L = 100$).  Once we have our final $\mb{H}$, we can use this to create $B'_k$ as follows:

\begin{equation}
\mb{S_{B'_k}} = \mb{S_{A'_k}} \mb {H}
\end{equation}

\begin{figure}[!h]
	\centering	\includegraphics[width=\columnwidth]{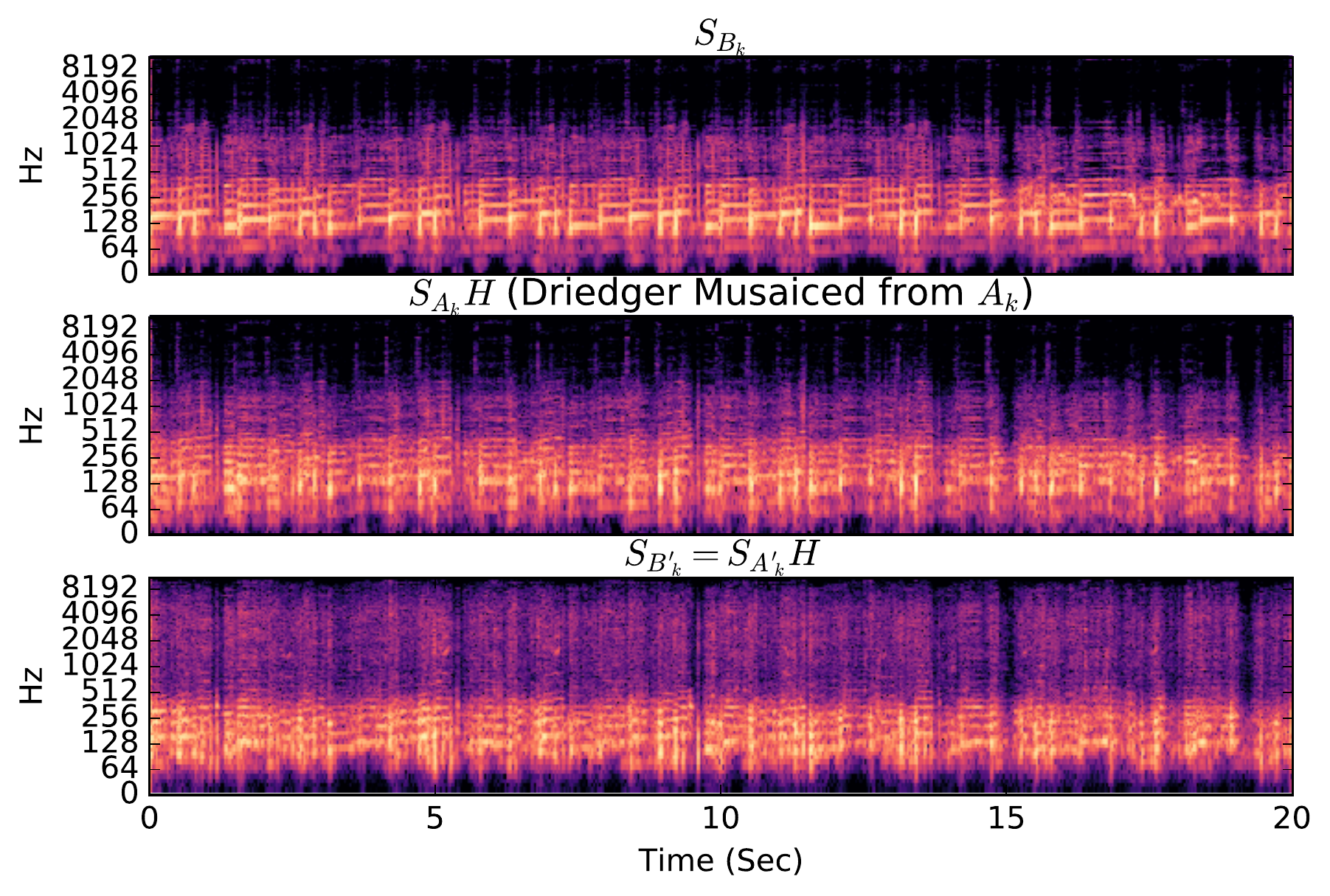}
	\caption{Driedger's technique \cite{driedger2015let} using audio grains from pitch shifted versions of $A_k$ (the $k^{\text{th}}$ track in Michael Jackson's ``Smooth Criminal'') to create $B_k$ (the $k^{\text{th}}$ track in Michael Jackson's ``Bad''), and using the activations to create $B'_k$ (the $k^{\text{th}}$ synthesized track in Alien Ant Farm's ``Bad'').}
    \label{fig:Driedger}
\end{figure}

In other words, we use the learned activations to create $\mb{S_{B_k}}$ using $\mb{S_{A_k}}$, but we instead use these activations with the dictionary from $\mb{S_{A'k}}$.  {\em This is the key step in the style transfer}, and it is done for the same reason that $\mb{H}_1$ is shared between $A$ and $A'$.  Figure~\ref{fig:Driedger} shows an example for the guitar track on Michael Jackson's ``Bad,'' translating to Alien Ant Farm using $A$ and $A'$ templates from Michael Jackson's and Alien Ant Farms' versions of ``Smooth Criminal,'' respectively.  It is visually apparent that $|\mb{S_{A_k}}| \mb {H} \approx \mb{S_{B_k}}$, it is also apparent that $\mb{S_{B'_k}} = \mb{S_{A'_k}} \mb {H}$ is similar to $\mb{S_{B_k}}$, except it has more energy in the higher frequencies.  This is consistent with the fact that Alien Ant Farm uses a more distorted electric guitar, which has more broadband energy.

To create our final $B'$, we simply add all of the STFTs $\mb{S_{B'_k}}$ together for each $k$, and we perform and inverse STFT to go back to audio.

\subsection{A Note About Tempos}
The algorithm we have described so far assumes that the tempos $t_A, t_{A'}$, and $t_B$ of $A$, $A'$, and $B$, respectively are similar.  This is certainly not true in more interesting covers.  Section~\ref{sec:Alignment} took care of the disparity between $A$ and $A'$ during the synchronization.  However, we also need to perform a tempo scaling on $B$ by a factor of $t_A/t_B$ before running our algorithm.  Once we have computed $B'$, whose tempo is initially $t_A$, we scale its tempo back by $(t_B/t_A) \cdot (t_{A'}/t_A)$.  For instance, suppose that $t_A = 60$, $t_{A'} = 80$, and $t_B = 120$.  Then the final tempo of $B'$ will be $60 \times (120/60) \times (80/60) = 160$ bpm.

\section{Experimental Examples}
\label{sec:experiments}

We now qualitatively explore our technique on several examples\footnote{Please listen to our results at \url{http://www.covers1000.net/analogies.html}}.  In all of our examples, we use $K=3$ sources.  This is a hyperparameter that should be chosen with care in general, since we would like each component to correspond to a single instrument or group of related instruments.  However, our initial examples are simple enough that we expect basically a harmonic source, a percussive source, and one other source or sub-separation between either harmonic and percussive.

First, we follow through on the example we have been exploring and we synthesize Alien Ant Farm's ``Bad'' from Michael Jackson's ``Bad'' ($B$), using ``Smooth Criminal'' as an example.  The translation of the guitar from synth to electric is clearly audible in the final result.  Furthermore, a track which was exclusively drums in $A$ included some extra screams in $A'$ that Alien Ant Farm performed as embellishments.  These embellishments transferred over to $B'$ in the ``Bad'' example, further reinforcing Alien Ant Farm's style.  Note that these screams would not have been preserved had we simply used inverted the CQT $\mb{\Lambda_{W_2, H_2}}$, but they are present in one of the filtered tracks and available as audio grains during musaicing for that track.

In addition to the ``Bad'' example, we also synthesize Alien Ant Farm's version of ``Wanna Be Startin Something,'' using ``Smooth Criminal'' as an example for $A$ and $A'$ once again.  In this example, Alien Ant Farm's screams occur consistently with the fast drum beat every measure.

Finally, we explore an example with a more extreme tempo shift between $A$ and $A'$ ($t_{A'} < t_A$).  We use Marilyn Manson's cover of ``Sweet Dreams'' by the Eurythmics to synthesize a Marilyn Manson cover of ``Who's That Girl'' by the Eurythmics.  We found that in this particular example, we obtained better results when we performed 2DNMF on $\mb{|C_{A}|}$ by itself first, and then we performed the optimizations in Equation~\ref{eq:updateW2} and Equation~\ref{eq:updateH}, holding $\mb{W_1}$ fixed.  This is a minor tweak that can be left to the discretion of the user at runtime.

Overall, our technique works well for instrument translation in these three examples.  However, we did notice that the vocals did not carry over at all in any of them.  This is to be expected, since singing voice separation often assumes that the instruments are low rank and the voice is high rank \cite{sprechmann2012real}, and our filters and final mosaicing are both derived from low rank NMF models.

\section{Discussion / Future Directions}

In this work, we demonstrated a proof of concept, fully automated end to end system which can synthesize a cover song snippet of a song $B$ given an example cover $A$ and $A'$, where $A$ is by the same artist as $B$, and $B'$ should sound like $B$ but in the style of $A'$.  We showed some promising initial results on a few examples, which is, to our knowledge, one of the first steps in the challenging direction of automatic polyphonic audio musaicing.  

Our technique does have some limitations, however.  Since we use $\mb{W_1}$ for both $A$ and $B$, we are limited to examples in which $A$ and $B$ have similar instruments.  It would be interesting to explore how far one could push this technique with different instruments between $A$ and $B$, which happens quite often even within a corpus by the same artist.

We have also noticed that in addition to singing voice, other ``high rank'' instruments, such as the fiddle, cannot be properly translated.  We believe that that translating such instruments and voices would be an interesting and challenging future direction of research, and it would likely need a completely different approach to the one we presented here.

Finally, out of the three main steps of our pipeline, synchronization (Section~\ref{sec:Alignment}), blind joint factorization/source separation (Section~\ref{sec:NMF2D}), and filtering/musaicing (Section~\ref{sec:musaicing}), the weakest step by far is the blind source separation.  The single channel source separation problem is still far from solved in general even without the complication of cover songs, so that will likely remain the weakest step for some time.  If one has access to the unmixed studio tracks for $A$, $A'$, and $B$, though, that step can be entirely circumvented; the algorithm would remain the same, and one would expect higher quality results.  Unfortunately, such tracks are difficult to find in general for those who do not work in a music studio, which is why blind source separation also remains an important problem in its own right.

\section{Acknowledgements}
Christopher Tralie was partially supported by an NSF big data grant DKA-1447491 and an NSF Research Training Grant NSF-DMS 1045133.  We also thank Brian McFee for helpful discussions about invertible CQTs.

\bibliographystyle{plain}
\bibliography{main}

%
%
%
%

\end{document}